# Chatty Mobiles:
# Individual mobility and communication patterns

Thomas Couronné, Zbigniew Smoreda, Ana-Maria Olteanu

Sociology and Economics of Networks and Services department
Orange Labs R&D, Paris

**Introduction**

Human mobility analysis is an important issue in social sciences, and mobility data are among the most sought-after sources of information in urban studies, geography, transportation and territory management. In network sciences mobility studies have become popular in the past few years, especially using mobile phone location data [1,2,3,4,5]. For preserving the customer privacy, datasets furnished by telecom operators are anonymized. At the same time, the large size of datasets often makes the task of calculating all observed trajectories very difficult and time-consuming. One solution could be to sample users. However, the fact of not having information about the mobile user makes the sampling delicate. Some researchers [1] select randomly a sample of users from their dataset. Others try to optimize this method, for example, taking into account only users with a certain number or frequency of locations recorded [2,3]. At the first glance, the second choice seems to be more efficient: having more individual traces makes the analysis more precise. However, the most frequently used CDR data (Call Detail Records) have location generated only at the moment of communication (call, SMS, data connection). Due to this fact, users' mobility patterns cannot be precisely built upon their communication patterns. Hence, these data have evident shortcomings both in terms of spatial and temporal scale.

In this paper we propose to estimate the correlation between the user's communication and mobility in order to better assess the bias of frequency based sampling. Using technical GSM network data (including communication but also independent mobility records - as in [6]), we will analyze the relationship between communication and mobility patterns.

**Data**

One weekday GSM data of the Paris Region territory (12,012 km² - 4,638 sq mi) were used. The dataset covers 4 million of French mobile phone users and more than 94 million records. Data are issued from Orange GSM network probes, they are anonymous (a secure, random network attributed temporary identity) and contain both cell localized communication events (calls and SMS) and mobility events (handover (HO) and location area update (LAU)).

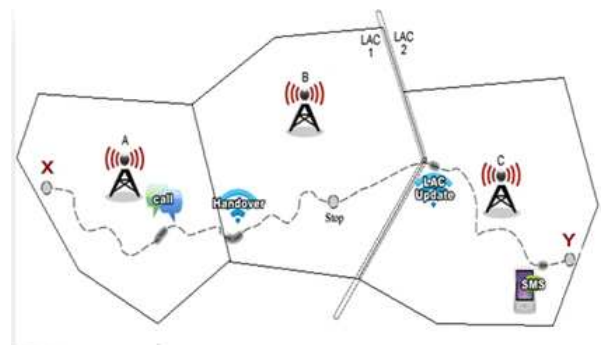

Figure 1. An example of mobile phone network localization data types for one user travelling from X to Y

Concerning mobility records, there are two types of data: HO data are generated during a communication, when a mobile phone changes position and is transferred to a new antenna; LAU records are generated when a device changes location area (in Paris Region, a location area groups on average 150 cells). The LAU is generated also when a mobile moves from one location area to the next while not on a call. It is exactly the information we need for our analysis as it is independent of the user communication behavior.

**Results**

Our analysis was conducted using two separate record types: communication data and communication-independent itinerary data. Communication frequency was plotted against the median number of mobility records (LAU) for each user (see: figure 2).

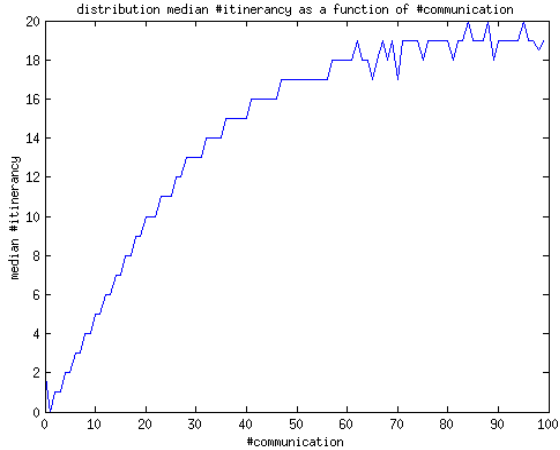

Figure 2. Median number of local area change as a function of communication frequency

Ninety percent of users have less than 30 communication events (calls or SMS) during the observed day. For this group, we notice a clear, almost linear correlation between the frequency of communications and the median number of location area changes (daily mobility indicator).
The curve reaches a plateau at about 50 communications per day and then the communication - mobility link disappears. People who communicate extremely frequently can no longer be distinguished by their median mobility.
The relationship between the number of itinerancy events (LAU) and the median communication frequency has also been studied. Again this correlation is nearly a perfect one: the more mobility records are, the more frequent mobile communications are.

To analyze conjointly user's mobility and communication, we constructed 8 equal frequency ordered classes for each event type (where class 1 is the lowest mobility/communication, and the class 8 is the most intensive itinerary/communication). Then we populated an 8x8 matrix with users having all specific combinations of itinerancy and communication events. The result of this operation is showed in figure 3.

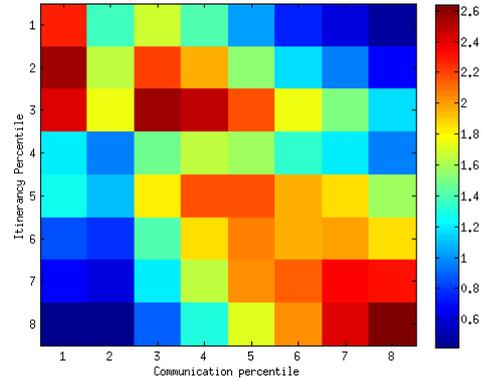

Figure 3. Conjoint distribution of users into communication and itinerancy intensity classes (1-low, 8-high)

The red color signifies the most probable combination and the deep blue color designates the less probable one. As we can observe, red squares are distributed on the top-right (both infrequent communication and mobility), on the bottom-left (both intense communication and mobility) as well as in the center of the matrix (average values). All other combinations, and in particular high-low combinations, are less frequent. The matrix indicates that in our data the relationship between user mobility and user communication frequency is really strong.

To complete this approach, the daily displacement distance per user was calculated using all localized records (calls, SMS, HO, and LAU) and compared to communication events distribution (see: figure 4).

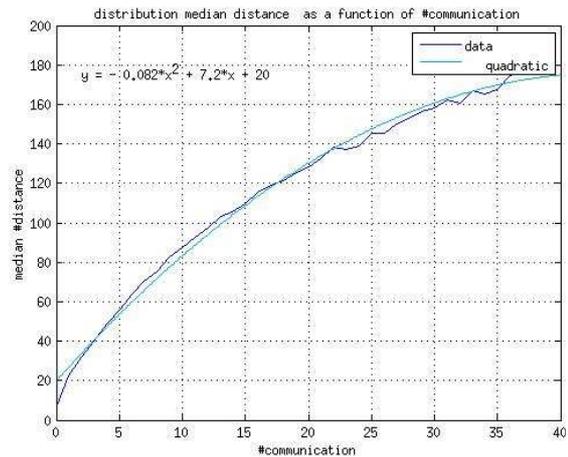

Figure 4. Median daily distance traveled (km) as a function of communication events frequency

We looked for a best regression model to fit our data, where $y$ is the median daily distance in km and $x$ is the number of communication events (call, SMS). It appears that the best model is a quadratic function:

$$y = -0.82x^2 + 7.2x + 20$$

This analysis confirms our observation showed in figure 2: the higher the number of communication events, the less strong the increase of the mobility distance.

**Conclusion**

A significant correlation between user mobility events and communication frequencies confirms our intuition that in mobile phone usages both phenomena are interrelated. A highly mobile person has in fact greater probability to use a mobile phone than someone who only commutes between a few places where s/he can also communicate *via* a fixed telephone. As his/her correspondents learn with time which is the most adapted communication channel to reach this person, they will also contribute to reinforce the observed correlation.

From the point of view of human mobility analysis using data from mobile phone, such as CDRs, our results ask for a very careful examination of sampling methods which are used. Selecting users with frequent communication traces, i.e., with many cell localizations, seems to introduce a clear bias because people having more mobile communications are also in a more mobile class of the general population.

Definitely, to better calibrate mobile phone data analysis in this domain a cross-analysis of mobile phone and survey-like data is needed.